\documentclass[iop]{emulateapj}
\usepackage{amsmath}
\usepackage{natbib}
\usepackage{pstricks}
\usepackage{mathrsfs}
\begin{document}

\newcommand{\uA}{\mathrm{A}}
\newcommand{\uM}{\mathrm{M}}
\newcommand{\ud}{\mathrm{d}}

\title{Determining the optimal locations for shock acceleration in magnetohydrodynamical jets}

\author{Peter Polko\altaffilmark{1}, David L. Meier\altaffilmark{2}, \& Sera Markoff\altaffilmark{1}}

\altaffiltext{1}{Astronomical Institute `Anton Pannekoek', University of Amsterdam, P.O. Box 94249, 1090 GE Amsterdam, the Netherlands; P.Polko@uva.nl}
\altaffiltext{2}{Jet Propulsion Laboratory, California Institute of Technology, Pasadena, CA 91109, USA}

\begin{abstract}
Observations of relativistic jets from black holes systems suggest that particle acceleration often occurs at fixed locations within the flow. These sites could be associated with critical points that allow the formation of standing shock regions, such as the magnetosonic modified fast point. Using the self-similar formulation of special relativistic magnetohydrodynamics by \citeauthor{2003ApJ...596.1080V}, we derive a new class of flow solutions that are both relativistic and cross the modified fast point at a finite height. Our solutions span a range of Lorentz factors up to at least 10, appropriate for most jets in X-ray binaries and active galactic nuclei, and a range in injected particle internal energy. A broad range of solutions exists, which will allow the eventual matching of these scale-free models to physical boundary conditions in the analysis of observed sources.
\end{abstract}


\section{Introduction}

Jets have been observed around a large variety of astrophysical objects, such as young stellar objects (YSOs), accreting white dwarfs, X-ray binaries (XRBs), and active galactic nuclei (AGN), and are also thought to drive gamma-ray bursts (GRBs). In YSOs jets facilitate angular momentum transport, allowing the central star to accrete more matter, and likely play a similar role in accreting black hole systems, where in AGN they are thought to also affect the evolution of their host galaxy \citep[e.g.][]{2006MNRAS.368L..67B}. While there are enormous differences between the types and scales of objects around which jets can occur, the origins of jets seem to be remarkably similar, requiring the basic ingredients of infalling/collapsing, rotating matter, and magnetic fields.

Despite these seemingly simple initial conditions, there are currently many outstanding problems in our understanding of jets, from their creation to their matter content and internal physics. One important facet of jets observationally is their hallmark synchrotron emission dominating the radio bands in particular, extending up to at least the near infrared (NIR) in XRBs \citep{1997MNRAS.290L..65F}. At higher frequencies the typical power-law spectral energy distributions (SEDs) are generally interpreted as optically thin synchrotron emission from accelerated particles, both in AGN \citep{1985ApJ...298..114M} and in XRBs \citep{2002LNP...589..101F}. An effective way to accelerate radiating particles into a power-law distribution is via diffusive shock acceleration \citep[e.g.][]{1978MNRAS.182..147B,1983RPPh...46..973D} off scattering centers in turbulent plasma flows. Once initiated, this process must to be distributed throughout the flow to account for the lack of spectral aging over vast distances along the jets \citep[e.g.][]{2001A&A...373..447J}.

But where does the acceleration itself begin, and what triggers it? There is increasing evidence that the start of this region is offset from the central compact object. For example in the jet of the AGN M87, recent VLBI observations show the synchrotron emission starting at a region offset by $\sim100r_g$ from the core \citep{1999Natur.401..891J,2008JPhCS.131a2053W}. Similarly, the start of power-law acceleration in compact jets would be indicated by a transition from optically thick emission, with a flat/inverted spectral index, to an optically thin power-law at a distinct location in the SED. Such a break has been observed directly so far only in one source, the Galactic XRB GX 339-4, in the NIR \citep{2002ApJ...573L..35C}, during the ``hard'' or ``nonthermally dominated'' accretion state associated with compact jet formation \citep[see state definitions in, e.g.,][]{2006csxs.book..157M}. Because of the stratification of emission regions in compact jets \citep[e.g.][]{1979ApJ...232...34B}, the lower the frequency where this turnover occurs, the further the location along the jet where particle acceleration starts. In XRBs, a break in the NIR corresponds to an offset of $\sim10-1000$ $r_g$, and models of the broadband data of most black hole XRBs in the hard state so far seem to require such a break \citep[e.g.][]{2001A&A...372L..25M,2005ApJ...635.1203M,2005ApJ...626.1006N,2007ApJ...670..610M,2009MNRAS.398.1638M}.

The fact that the start of the acceleration region seems to occur at roughly the same location in several systems could be indicative of a critical point occurring in a magnetohydrodynamical (MHD) flow, particularly the magnetosonic modified fast point (MFP). At the MFP the collimating magnetic field lines turn inwards towards the jet axis, potentially leading to recollimation shocks, while at the same time the flow becomes causally disconnected so shocks can occur without disrupting the flow upstream. Such a shock region thus would occur at a fixed location in the flow, closely connected to the MFP, and would be an ideal location for particle acceleration to begin. We wish to investigate the feasiblity of this premiss in this paper.

Because of the complexity involved in relativistic MHD including strong gravity, many groups are using the results of simulations to study the formation and development of jets. These simulations often show the development of a steady outflow in which the magnetic field is remarkably self-similar and axisymmetric near the launch point \citep[see, e.g., Fig. 2 and Fig. 11 in][]{2006MNRAS.368.1561M}. However to study specifically the development of critical points in the flow and their dependence on external boundary conditions, current MHD simulations either do not extend far enough from the black hole, or if they do, they are too computationally expensive. Assuming that the resulting flows retain a self-similar structure, at least when gravity does not dominate as indicated in the simulations, we adopt the formalism developed by \citet[hereafter VK03]{2003ApJ...596.1080V}. By assuming axisymmetry and a self-similar field line geometry, VK03 reduce the exact equations of special relativistic MHD to a one-dimensional problem. Although VK03 focused on jets in GRBs, this treatment is also applicable to other MHD jets such as in AGN and microquasars.

In an earlier self-similar treatment that was non-relativistic, \citet{2000MNRAS.318..417V} presented a solution where the flow crosses the MFP at a finite height above the disk. In VK03, however, they only found relativistic solutions with an MFP occurring at infinity (meaning that the flow asymptotically approaches a perfect cylindrical geometry). In this paper we extend the study of VK03 and derive new solutions in which the relativistic flow crosses the MFP at a finite location above the disk. In \S~2 we describe the VK03 model and our method for exploring the full parameter space of solutions. In \S~3 we present the first relativistic solutions that pass through an MFP. In \S~4 we discuss our results, and the dependence of the MFP location on the model parameters. We also describe how this work sets the stage for further development to connect the flow to regions near the disk where gravity can no longer be ignored. \S~5 contains our conclusions.

\section{Method}
\subsection{Background}

During the acceleration and collimation of jets, magnetic fields are thought to efficiently extract rotational energy from either the compact object \citep{1977MNRAS.179..433B} or the accretion disk \citep{1982MNRAS.199..883B}. The latter models are in the Newtonian limit, with the matter considered cold, meaning there is negligible thermal pressure causing bulk acceleration to non-relativistic velocities.

These cold, non-relativistic solutions were generalized to the relativistic regime by \citet{1992ApJ...394..459L}, allowing the bulk velocity to attain relativistic speeds. VK03 further extended the solutions to include the ``hot'' regime, allowing the random motions of the particles to become relativistic and thus the jets to be hydrodynamically accelerated even at the base where temperatures are high. It is this last scenario that we base this work upon.

\begin{figure}
\includegraphics[width = 0.45 \textwidth]{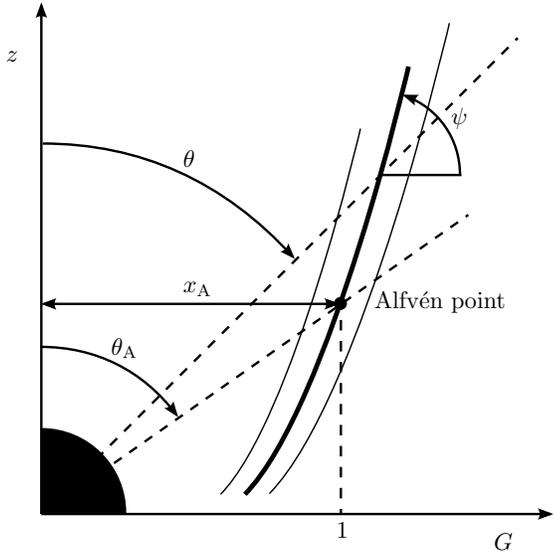}
\caption{Sketch of self-similar field lines projected onto the meridional plane. For any $\theta$ the values of the variables describing the field line are exactly the same, only scaled by their respective distances.}
\label{geometry}
\end{figure}

Starting from the equations of time-dependent special relativistic MHD, we make the following assumptions to render them more tractable: ideal MHD, no gravitational field or external force (and thus self-similarity holds), axisymmetry, a zero azimuthal electric field and time independence. Following the terminology of VK03, after scaling the equations to make them non-dimensional, we are left with two coupled differential equations (equations \eqref{G} and \eqref{transfield} in the appendix). Combining these two coupled differential equations, we obtain a single equation for $\ud M^2/\ud \theta$ (equation \eqref{windequation}, with $M$ the Alfv\'enic Mach number and $\theta$ the angle of the point on the field line with the axis of symmetry) which acts like a ``wind equation'', much akin to the wind equation of the Parker solar wind model \citep{1958ApJ...128..664P}. This wind equation, along with the other algebraic equations (see appendix), can be solved for the velocity, magnetic and electric field strength, density and pressure along a field line. Due to the self-similar assumption, once solutions are obtained for one field line, all other field lines can be obtained by simple scaling. An example of this self-similarity and the meaning of some of the parameters used can be found in Figure \ref{geometry}.

Instead of the single critical (or sonic) point of the Parker solar wind model, due to the inclusion of magnetic fields, the obtained wind equation has three locations where the denominator crosses zero. Starting from the accretion disk, these are the modified slow point (MSP), the Alfv\'en point, and the MFP. The MSP and MFP are also called the slow and fast magnetosonic separatrix surfaces. The Alfv\'en point is the location where the relativistic collimation speed of the flow towards the axis ($V_\theta$) is given by
\begin{equation}
\left( \gamma V_\theta \right)^2 = \frac{B_\theta^2 \left( 1 - x^2 \right)}{4 \pi \rho_0 \xi},
\end{equation}
where $\gamma = 1/(1 - V^2/c^2)^{1/2}$ is the Lorentz factor, $B$ the strength of the magnetic field, $x$ the cylindrical radius in units of the light cylinder radius, $\rho_0$ the baryon rest-mass density, and $\xi c^2$ the specific relativistic enthalpy (the variables are described in more detail in \S \ref{variables}). The denominator of $\ud M^2/\ud \theta$ with the Alfv\'en point divided out, can be expressed as
\begin{align}
\mathscr{D} =& \left( \frac{\gamma V_\theta}{c} \right)^4 - \left( \frac{\gamma V_\theta}{c} \right)^2 \left[ \frac{U_\mathrm{s}^2}{c^2} + \frac{B^2 - E^2}{4 \pi \rho_0 \xi c^2} \right] \nonumber \\
&+ \frac{U_\mathrm{s}^2}{c^2} \frac{B_\theta^2 (1 - x^2)}{4 \pi \rho_0 \xi c^2},
\end{align}
with
\begin{equation}
U_\mathrm{s}^2 = c^2 \frac{(\Gamma - 1) (\xi - 1)}{(2 - \Gamma) \xi + \Gamma -1},
\end{equation}
$c$ the velocity of light, $E$ the strength of the electric field, and $\Gamma$ the polytropic index. The MSP and MFP are, by definition, the locations where $\mathscr{D} = 0$ (VK03).

At every critical point, the numerator of $\ud M^2/\ud \theta$ should also pass through zero to ensure a smooth crossing. This translates into a regularity condition at the critical points and fixes the value of a free parameter. Even though the MSP should be crossed smoothly to obtain a solution that describes the entire jet from the accretion disk to the termination point, gravitational effects cannot be ignored at the MSP. Since the equations do not include gravity (as it is not compatible with the self-similarity assumption in relativistic flow), we do not try to fit for the MSP. Therefore we fit two critical points, and correspondingly two parameters are fixed, in our approach described below $p_\uA$ for the Alfv\'en point and $\sigma_\uM$ for the MFP.

The physical importance of the MFP is that it is the location where not even the fastest signals can travel upstream anymore, meaning anything downstream from the MFP is causally disconnected from the region upstream. If, for example, a shock were to exist beyond the MFP, it could not disrupt the flow leading to that shock, allowing it to be a permanent feature. As mentioned above, at the MFP there is a component of the velocity heading towards the axis. This can lead to a collimation shock shortly beyond the MFP, causing the magnetic energy to be converted to particle energy and the jet to become kinetically dominated. Another possibility is that the flow remains magnetically dominated, and, after reaching a minimum radius, bounces back, retaining an ordered magnetic field \citep[see, e.g.,][]{1994ApJ...429..139C}. This is not in conflict with the statement in VK03 that the only physically acceptable case in the super-Alfv\'enic regime is for the flow to become asymptotically cylindrical, as this statement only applies to solutions with $F > 1$.

\subsection{Model parameters}
\label{variables}

The prescription we are following from VK03 has 9 free parameters that determine the solution, whose effects are described below. Following the same notation, a Roman subscript A signifies the value of a variable at the Alfv\'en point and an Italic {\it A} denotes a value with respect to the poloidal magnetic flux function.

\subsubsection{Free parameters}

The exponent $F$ determines the current distribution. A value $F > 1$ corresponds to the current-carrying regime, with higher values of F ensuring faster collimation, but if $F < 1$ we are in the return-current regime. The restriction on $F$ is that it cannot be negative: \mbox{$F > 0$}. Although we consider $F$ to be a free parameters, for this paper we chose to keep it fixed at 0.75.

The adiabatic index $\Gamma$ can have values of $4/3$ for relativistic and $5/3$ for non-relativistic solutions.

$\theta_\uA$ gives the angle where the Alfv\'en point is located with respect to the axis of symmetry. It is limited by the value of $\psi_\uA$: \mbox{$\theta \in \langle 90^\circ - \psi_\uA, 90^\circ]$}.

$\psi_\uA$ gives the poloidal slope of the field line at the Alfv\'en point with respect to the accretion disk. This in turn is limited by $\theta_\uA$: \mbox{$\psi_\uA \in \langle 90^\circ - \theta_\uA, 90^\circ]$}.

$x_\uA^2$ is the radius squared of the Alfv\'en point in terms of the light cylinder radius. For $x_\uA^2 \rightarrow 1$ the solution becomes more force-free. The allowed values are \mbox{$x_\uA^2 \in \langle 0, 1\rangle$}.

$\sigma_\uM$ is the magnetization parameter in the monopole solution of \citet{1969ApJ...158..727M} and is related to the mass-to-magnetic flux ratio. The constraint is \mbox{$\sigma_\uM > 0$}.

$q$ is the dimensionless adiabatic coefficient and is constant along a field line. For a large value of $q$, the specific relativistic enthalpy of the matter is high, for $q \rightarrow 0$ equation \eqref{M2} shows $\xi \rightarrow 1$ and the flow is cold. Therefore \mbox{$q \ge 0$}.

$p_\uA$ is the derivative of $M^2$ with respect to the polar angle $\theta$ at the Alfv\'en point. Accelerating flow implies \mbox{$p_\uA < 0$}.

$B_0 \varpi_0^{2-F}$, with reference magnetic field $B_0$ and reference length $\varpi_0$, is the scaling of the solution, relating the dimensionless values to physical dimensions. We do not yet apply our solutions to specific black hole systems, so this parameter is not used here but it will be important for future applications of our solutions.

\begin{figure}
\includegraphics[width = 0.5 \textwidth]{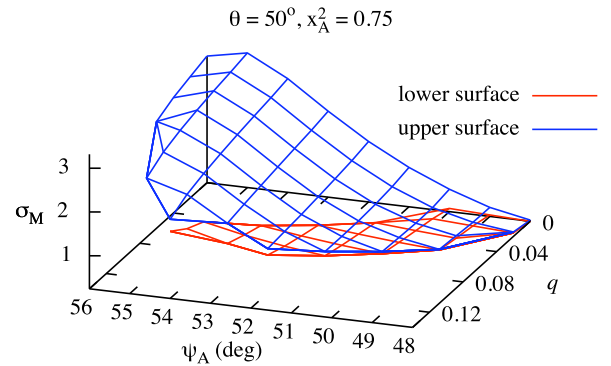}
\caption{3D plot of solutions with an MFP with parameters $F = 0.75$, $\Gamma = 5/3$, $x_\uA^2 = 0.75$ and $\theta_\uA = 50^{\circ}$. On the front side the solid blue and dashed red surfaces are connected. The lower surface does not extend all the way to the $\psi_\uA$-axis for all values of $\psi_\uA$. Because we favor solutions with strong magnetic fields, we focus on the upper surface, with high $\sigma_\uM$.\\
(A color version of this figure is available in the online journal.)}
\label{3d}
\end{figure}

The smooth crossing of the Alfv\'en point is ensured by calculating $p_\uA$ from the corresponding regularity condition, given by equation \eqref{ARC}. In the same way we determine $\sigma_\uM$ by crossing the MFP, although this parameter sometimes can have two values (see Figure \ref{3d}). This leaves $F$, $\Gamma$, $\theta_\uA$, $\psi_\uA$, $x_\uA^2$, and $q$ (and $B_0 \varpi_0^{2-F}$) to satisfy the boundary conditions at the source and, indirectly, at the end of the jet.

\subsubsection{Other parameters and variables}

There are also parameters derived from the above values:

$\mu$ determines the total energy-to-mass-flux ratio ($\mu c^2$) and is conserved along a field line. This parameter is determined from equation \eqref{mu2} and \mbox{$\mu > 1$}.

$\sigma_\uA$ is the value of the magnetization function $\sigma$, defined as the Poynting-to-matter energy flux, at the Alfv\'en point. This parameter is determined from equation \eqref{sigma_A}. As $\mu$ cannot be negative, from equation \eqref{mu2} follows \mbox{$\sigma_\uA \in [ 0, \frac{x_\uA}{1 - x_\uA} ]$}.

And finally we describe the other variables used in the equations:

$\xi$ determines the specific (per baryon mass) relativistic enthalpy ($\xi c^2$). If $\xi = 1$ we have cold, pressureless matter. $\xi$ will drop from the high temperatures at the base of the jet as matter is mainly accelerated hydrodynamically and at some point above the disk drop down to 1. From this point on all acceleration is magnetic. This variable is determined from equation \eqref{M2}.

$M$ is the Alfv\'enic Mach number, the velocity of the flow in terms of the Alfv\'en velocity.

$G$ is the radius in terms of $x_\uA$ and is therefore equal to 1 at the Alfv\'en point.

\subsection{Numerical method}

\begin{figure}
\includegraphics[width = 0.45 \textwidth]{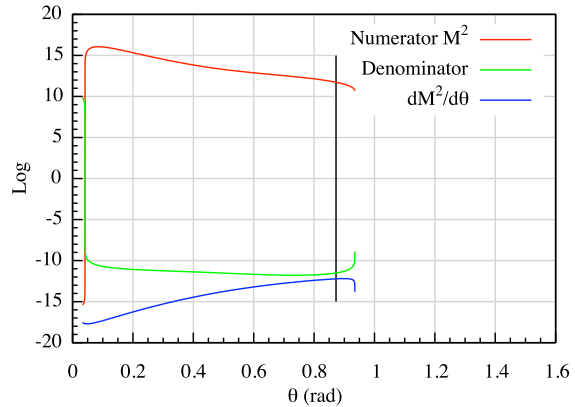}
\caption{Solution c. For parameters see Table \ref{parametertable}. The scaling for the $y$ axis is a combination of a linear and logarithmic part, using the function $\mathrm{sign}(x) \log_{10}[1 + \mathrm{abs}(x)/10^{-12}]$. The vertical black line gives the location of the Alfv\'en point (at $\theta \approx 0.87$). Even though the numerator (solid red line) and denominator (dashed green line) change sign, it can be seen from their ratio (short dashed blue line, $\mathrm{d} M^2/\mathrm{d} \theta$) that the MFP is crossed smoothly. We ceased the integration shortly after the MFP.\\
(A color version of this figure is available in the online journal.)}
\label{solutionc}
\end{figure}

To find solutions with an MFP we obtained expressions for $\ud M^2 / \ud \theta$, given by equation \eqref{windequation}, and for $\ud \psi / \ud \theta$ by combining the derivative of the Bernoulli equation with the transfield equation using the determinant method. Because $\ud \psi / \ud \theta$ is very unstable near the Alfv\'en point, as the numerator has a first order zero point there and the denominator a second order one, we reverted to the Bernoulli equation, given by \eqref{Bernoulli}, to determine $\psi$. To start off the integration from the Alfv\'en point we specify $F$, $\Gamma$, $\theta_\uA$, $x_\uA^2$, $\psi_\uA$, $q$ and an initial guess for $\sigma_\uM$. We determine $p_\uA$ from the Alfv\'en regularity condition, equation \eqref{ARC}. Integrating outward from the Alfv\'en point, we determine whether the numerator or denominator crosses zero first and adjust $\sigma_\uM$ accordingly until both cross at the same time. We then proceed to explore the range of solutions which cross both the Alfv\'en point and MFP (see Figure \ref{solutionc}). For plotting purposes, we have divided out the factor $x^6 (1 - M^2 - x^2)^2$ in both the numerator and denominator.

As discussed above, because there is no gravity in the model, we do not try explicitly to cross the modified slow point (MSP). Gravitational effects should play a large role close to the black hole, and the self-similar equations cannot predict accurately where the MSP is located.

\subsection{Approach}

To begin our exploration of parameter space, we chose a solution from \citet{2000MNRAS.318..417V} known to have an MFP (specified below their Figure 4) with parameters $x = 0.75$, $\gamma = 5/3$, $\theta_* = 60^{\circ}$, $\psi_* = 45^{\circ}$, $\kappa^2 = 15$, $\mu = 10.9239$, $\lambda^2 = 2.7935$, $p_* = -5.5744$ and $\varepsilon = 9.4487$. By comparing terms in \citet{2000MNRAS.318..417V} and VK03, it is possible to translate these parameters to the parameters used in VK03. Because we are using the relativistic equations from VK03, the parameters of our first solution (see Table \ref{parametertable}) differ slightly from the corresponding parameters above and we vary $\sigma_\uM$ to obtain a critical solution again. From this solution we were able to traverse parameter space, while allowing only critical solutions with an MFP. By increasing $x_\uA^2$ we were able to obtain higher velocities of the jet. After achieving relativistic velocities, we focused on finding solutions with higher values of $q$, as the solutions so far were cold. But for a fixed value of $x_\uA^2$ there is a maximum value of $q$ that produces critical solutions crossing the MFP. This is due to the fact that the collection of all solutions form a surface in the multidimensional parameter space, and we had reached a maximum for $q$ for the fixed values of the other parameters (see Figure \ref{3d}).

The surfaces of valid solutions have roughly the same appearance for the explored range of $\theta_\uA$ and $x_\uA^2$. To describe the effect of $\theta_\uA$ and $x_\uA^2$ on the solutions, we approximate the graph as a cone with the base in the $\psi_\uA,\sigma_\uM$-plane and with the maximally allowed value of $q$ as its height. If we increase $x_\uA$ the base of the cone shrinks while moving to higher $\sigma_\uM$, and the height decreases. The latter limits the maximum value that $x_\uA^2$ can be increased to. If we decrease $\theta_\uA$, the height increases, indirectly allowing higher values for $x_\uA^2$. The area of the base becomes bigger and shifts to higher $\psi_\uA$ and $\sigma_\uM$, with the upper surface becoming steeper. Due to the shape of the surface, it is possible to increase any two parameters of $x_\uA^2$, $q$ and $\sigma_\uM$ at the expense of the third.

By extending our search to three parameters ($\psi_\uA$, $q$ and $\sigma_\uM$), we were able to move around this point and continue increasing $q$. This revealed a multidimensional surface that is double-valued in $\sigma_\uM$. An example of this surface is shown in Figure \ref{3d}, which also includes our most relativistic solution presented below, solution $c$. The numerator, denominator, and acceleration of $M^2$ in this latter solution are also plotted in Figure \ref{solutionc}.

\section{Results}

\begin{figure*}
\includegraphics[width = 0.95 \textwidth]{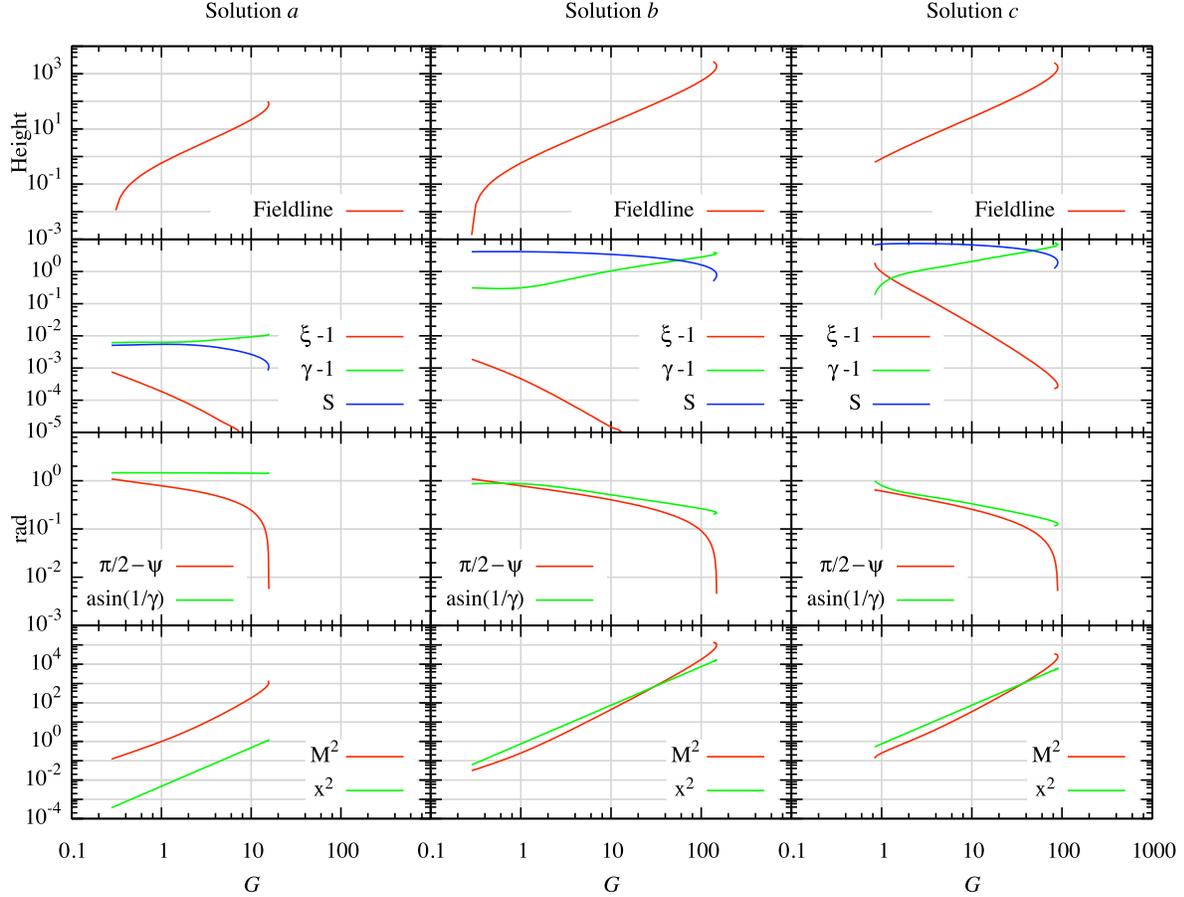}
\caption{Properties of solutions. Row 1 shows the geometry of the field line, where the height has the same scaling as the cylindrical radius. Row 2 shows the Poynting-to-mass flux ratio ($S \equiv - \varpi \Omega B_\phi / \Psi_\uA c^2$), the Lorentz factor $\gamma$, and the specific relativistic enthalpy. Row 3 shows the causal connection opening angle arcsin($1/\gamma$), and the opening half-angle of the outflow ($\pi/2 - \psi$), which is negative after overcollimation. Row 4 shows the squares of the Alfv\'enic Mach number ($M^2$), and the radius in units of the light cylinder radius ($x^2$).\\
(A color version of this figure is available in the online journal.)}
\label{overview}
\end{figure*}

In this section we present the various solutions crossing the MFP, from the first one that we found to one with relativistic temperature and bulk flow we sought, while describing the features particular to a certain solution. The parameters of our solutions are given in Table \ref{parametertable} and the main properties in Figure \ref{overview}. As we are not yet applying our solutions to specific black hole systems, we do not use the scaling parameters $B_0 \varpi_0^{2-F}$ and $\varpi_0$.

\begin{table*}[htdp]
\caption{Parameters of solution}
\begin{center}
\begin{tabular}{l c c c c c c c c c c}
\hline
\hline
 & & & & & $\theta_\uA$ & $\psi_\uA$ & & & & \\
Solution & $F$ & $\Gamma$ & $x_\uA^2$ & $q$ & (deg) & (deg) & $\sigma_\uM$ & $p_\uA$ & $\sigma_\uA$ & $\mu$ \\
\hline
$a$ & 0.75 & 5/3 & $4.7676 \times 10^{-3}$ & $2.5348 \times 10^{-6}$ & 60 & 45 & $2.14228 \times 10^{-4}$ & -5.54314 & $5.42665 \times 10^{-3}$ & 1.01197 \\
$b$ & 0.75 & 5/3 & 0.75 & $2.5 \times 10^{-6}$ & 60 & 45 & 1.01241 & -1.50363 & 3.14708 & 5.44084 \\
$c$ & 0.75 & 5/3 & 0.75 & 0.12 & 50 & 55 & 2.53981 & -1.66651 & 2.6039 & 9.85117 \\
\hline
\end{tabular}\\
\end{center}
{\sc Note.}---The values for the first six parameters ($F$ through $\psi_\uA$) are exact, for the last four ($\sigma_\uM$ through $\mu$) they are rounded off.
\label{parametertable}
\end{table*}

\subsection{Solution $a$: A cold, slow jet}

This solution is the closest to the non-relativistic parameter values given in \citet{2000MNRAS.318..417V} that conforms to the Alfv\'en regularity condition (ARC), which is the transfield equation at the Alfv\'en point. The solution crosses the MFP at $\theta \approx 0.15$ or $8.6^{\circ}$ ($G \approx 15.3$). The Alfv\'en point is located at $\theta \approx 1.05$ or $60^{\circ}$. The top left panel of Figure \ref{overview} gives the meridional projection of the magnetic field lines. The jet overcollimates after a maximum radius of almost 16 times the Alfv\'en radius at $\theta \approx 0.20$ or $11.5^{\circ}$, shortly before the MFP. The second left panel of Figure \ref{overview} shows the flow is cold throughout ($\xi - 1 \ll 1$) due to the very small value for $q$ and low $x_\uA^2$. As $M^2 = 1 - x_\uA^2$ at the Alfv\'en point, equation \eqref{M2} gives a value for $\xi_\uA$ very close to 1. The energy of the matter $\xi$ (including the dominant rest mass energy) is much higher than the energy in the magnetic field throughout. Therefore, even though the magnetic acceleration is efficient, the jet is not accelerated to relativistic velocities ($\gamma < 1.02$). The third center panel shows the ``causal connection'' opening angle $\arcsin(1/\gamma)$ and the opening half-angle of the outflow, which goes from $60^{\circ}$ to a few degrees overcollimation. Although the causal connection opening angle has little importance for non-relativistic flows, as it remains very close to $90^{\circ}$ since $\gamma \sim 1$, it is shown for completeness.

\subsection{Solution $b$: A cold, fast jet}

After our first solution, we increased the velocity of our jet by increasing $x_\uA^2$. As the top center panel in Figure \ref{overview} shows, after a long period where the field line remains almost parabolic, the jet in this solution overcollimates as well. This is caused by magnetic hoop stresses and may allow a shock region to develop beyond the MFP. The Alfv\'en point is again located at $\theta \approx 1.05$ and the MFP at $\theta \approx 0.063$ or $3.6^{\circ}$ ($G \approx 145$). The second center panel shows the Lorentz factor and the enthalpy of the flow. The flow here also is seen to be cold ($\xi \approx 1$), meaning the jet is mainly magnetically accelerated, which is again due to the small value of $q$. The Poynting-to-mass flux ratio ($ S \equiv -\varpi \Omega B_\phi / \Psi_A c^2$) decreases, showing magnetic energy being transferred into kinetic energy, with the flow reaching a Lorentz factor of 2.8. The bottom center panel shows the squares of the light cylinder radius, $x \equiv {\varpi \Omega}/{c}$, and the Alfv\'enic Mach number, $M$. When $x^2 = 1$ the light surface is reached, which is the radius where the field circular velocity reaches the speed of light.

\subsection{Solution c: A warm, very fast jet}

After having achieved a relativistic solution for the cold plasma case, we would like to find solutions with an increased flow temperature. To do so requires increasing the value of $\xi$. As $M^2$ is given by $1 - x_\uA^2$ at the Alfv\'en point, equation \eqref{M2} shows that increasing $x_\uA^2$ and/or $q$ has the desired effect. Unfortunately, for larger $x_\uA^2$ the maximum value of $q$ decreases. By choosing a lower value for $\theta_\uA$ the attainable values for $x_\uA^2$ and $q$ are increased, leading to a warmer flow. This solution is shown in the third column of Figure \ref{overview}. The Alfv\'en point is located at $\theta \approx 0.87$ or $50^{\circ}$ and the MFP at $\theta \approx 0.041$ or $2.4^{\circ}$ ($G \approx 87.4$). Near the beginning of the flow $\xi \approx 2.9$.

The Lorentz factor of the flow at the MFP is 8.3, which is mainly due to the high initial Poynting flux.

It can be seen in the second right panel of Figure \ref{overview} that $\xi$ always drops to 1 and from the fourth right panel that $M^2$ always dominates $x^2$ near the MFP. As the Lorentz factor is given by
\begin{equation}
\gamma = \frac{\mu}{\xi}\frac{1 - M^2 - x_\uA^2}{1 - M^2 - x^2}
\end{equation}
the final Lorentz factor is approximately $\mu$. This means that while the jets may start as Poynting flux-dominated, eventually they convert most of their Poynting flux and become kinetic energy-dominated.

\subsection{Location of the MFP}

Since we are interested in the location of the MFP, we would like to know how it depends on the model parameters. As it is challenging to sample the full parameter space, we will focus on the region around the solution closest to observed jets, solution $c$. By allowing all parameters to vary and looking at the effect this has on the location where the MFP occurs, we can draw the following conclusions: the MFP moves outward (smaller $\theta$) when the Alfv\'en point occurs at a smaller angle, (lower $\theta_\uA$), when the temperature at the base of the flow is increased (higher $q$), when the flow at the Alfv\'en point is already close to collimation (higher $\psi_\uA$), or when the Alfv\'en point moves closer to the light cylinder radius, making the flow more force-free (higher $x_\uA$).

\section{Discussion}

We have succeeded in obtaining new, solutions for a relativistic, magnetized flow that smoothly crosses the MFP at a finite height ($\theta > 0$) above the system equator. These solutions suggest that it should be possible to construct better, more MHD-consistent jet models where the location of the acceleration region is determined \emph{a priori} from the physical boundary conditions.

So far none of the solutions derived remains Poynting flux-dominated up to the MFP, which is probably due to the relatively small value of $\sigma_\uM$ found so far. Increasing $x_\uA^2$ and especially decreasing $\theta_\uA$ will allow higher values of $\sigma_\uM$ to be used. The same can also be done at the expense of $q$.

We also have described some of the relations between the different parameters (see \S \ref{variables}) and the effect they have on each other. Having a more force-free solution (higher $x_\uA^2$) decreases the allowed range of temperatures ($q$) and collimation at the Alfv'en point ($\psi_\uA$), but at the same time allows a higher value of $\sigma_\uM$ that provides a critical solution. Keeping all other parameters fixed, there is a maximum value of $x_\uA^2$ for which a solution is possible at all. This maximum may be increased by moving the Alfv\'en point closer to the disk (smaller $\theta_\uA$). This change has the additional effects of allowing a broader range for the collimation angle at the Alfv\'en point ($\psi_\uA$) while at the same time shifting this range towards higher collimation. It also allows for a higher temperature of the flow ($q$) or a higher magnetic field strengths ($\sigma_\uM$). Any pair of parameters $x_\uA^2$, $q$ and $\sigma_\uM$ may be increased at the expense of the third.

Two parameters not varied so far are $F$ and $\Gamma$. Higher values for $F$ should ensure faster collimation and might therefore be very important for the exact location of the MFP. Similarly, we may want to explore the $\Gamma = 4/3$ case for jets with extremely relativistic temperature. However, for the weaker jets in AGN and XRBs that we plan to target, the radiating particle distributions are generally thought to peak at mildly relativistic energies.

\section{Conclusion}

If the start of the particle acceleration region in steady jets is indeed associated with the magnetosonic fast critical point in the bulk flow, then our results support the conclusion that such a region could occur at a fairly stable location about the launch point. All of the solutions found overcollimate shortly before the MFP, as would be expected for the initiation of shock development. By starting with a solution that crosses the MFP in the non-relativistic case of \citet{2000MNRAS.318..417V}, we were able to extend the solution through the multidimensional parameter space towards relativistic velocities and temperatures, while retaining the critical point. This feature sets our results apart from the work of VK03, whose formulism we adopted, in that they are immediately applicable to observed compact jet sources with an optically thick-to-thin break in the synchrotron spectrum, such as hard state XRBs and weakly accreting AGN. Our most promising solution is a jet outflow with mildly relativistic temperature, and Lorentz factor of $\sim 10$, also appropriate for the steady jets in both XRBs as well as AGN.

It is clear that a wide range of parameter space is left still unexplored, that can be exploited for matching physical boundary conditions appropriate to known astrophysical sources. However, before a radiative model can be constructed around the dynamical ``backbone'' provided by the solutions presented here, a prescription for including gravity must be included, to extend these solutions through the MSP and allow connection with a physical model of the accretion flow/corona. We are currently working on matching these necessarily non-self similar solutions to those presented here, which will be presented in a separate work. Once this solution is in place, we will have a much more physically consistent model \citep[compared to, e.g.,][]{2005ApJ...635.1203M} to use in the fitting of data from accreting black holes across the mass scale, which show compact, steady jets. As there seems to be no shortage of possible solutions, we are confident that we can match physical boundary conditions with critical solutions.

\acknowledgements
P. Polko and S. Markoff gratefully acknowledge support from a Netherlands Organization for Scientific Research (NWO) Vidi Fellowship. In addition, S. Markoff is grateful for support from the European Community's Seventh Framework Program (FP7/2007-2013) under grant agreement number ITN 215212 ``Black Hole Universe''. Part of the research described in this paper was carried out at the Jet Propulsion Laboratory, California Institute of Technology, under a contract with the National Aeronautics and Space Administration. We would like to thank the anonymous referee for helpful comments that improved this manuscript.

\appendix
\section{Equations}
Here we list the equations we have used for reference. See \S \ref{variables} for a description of the parameters and variables.

\begin{equation}
x = x_\uA G
\label{x}
\end{equation}

\begin{equation}
M^2 = q \frac{\xi}{(\xi - 1)^{1/(\Gamma - 1)}}
\label{M2}
\end{equation}

The Bernoulli equation is given by
\begin{equation}
\frac{\mu^2}{\xi^2} \frac{G^4 (1 - M^2 - x_\uA^2)^2 - x^2 (G^2 - M^2 - x^2)^2}{G^4 (1 - M^2 - x^2)^2} = 1 + \frac{F^2 \sigma_\uM^2 M^4 \sin^2(\theta)}{\xi^2 x^4 \cos^2(\psi + \theta)}
\label{Bernoulli}
\end{equation}

\begin{equation}
\frac{\ud G^2}{\ud \theta} = \frac{2 G^2 \cos(\psi)}{\sin(\theta) \cos(\psi + \theta)}
\label{G}
\end{equation}

The transfield equation is given by
\begin{align}
G \sin^2(\theta) \frac{\ud}{\ud \theta} \left[ \tan(\psi + \theta) \frac{1 - M^2 - x^2}{G} \right] &= (F-1) \frac{x_\uA^4 \mu^2 x^2}{F^2 \sigma_\uM^2} \left( \frac{1 - G^2}{1 - M^2 - x^2} \right)^2 \nonumber \\
& - \sin^2(\theta) \frac{M^2 + F x^2 - F + 1}{\cos^2(\psi + \theta)} \nonumber \\
& - \frac{x_\uA^4 \mu^2 x^2}{F^2 \sigma_\uM^2 M^2} \left( \frac{G^2 - M^2 - x^2}{1 - M^2 - x^2} \right)^2 \nonumber \\
& + 2 \frac{\Gamma - 1}{\Gamma} \frac{F - 2}{F^2 \sigma_\uM^2} \frac{\xi (\xi - 1) x^4}{M^2}
\label{transfield}
\end{align}
The magnetization function $\sigma$ and the fractions at the Alfv\'en point are given by
\begin{equation}
\begin{array}{cc}
\sigma_\uA = \frac{2 x_\uA^2 \cos(\psi_\uA)}{p_\uA \sin(\theta_\uA) \cos(\theta_\uA + \psi_\uA)}, & \left( \frac{1 - M^2 - x_\uA^2}{1 - M^2 -x^2} \right)_\uA = \frac{1}{\sigma_\uA + 1}, \\
\left( \frac{1 - G^2}{1 - M^2 -x^2} \right)_\uA = \frac{\sigma_\uA / x_\uA^2}{\sigma_\uA + 1}, & \left( \frac{G^2 - M^2 - x_\uA^2}{1 - M^2 -x^2} \right)_\uA = \frac{x_\uA^2 - (1 - x_\uA^2) \sigma_\uA}{x_\uA^2 (\sigma_\uA + 1)}.
\label{sigma_A}
\end{array}
\end{equation}
We obtain $\mu^2$ by inserting these relations into the Bernoulli equation.
\begin{equation}
\mu^2 = \frac{(\sigma_\uA+1)^2}{x_\uA^2 - \left[x_\uA^2 - \sigma_\uA \left(1 - x_\uA^2 \right) \right]^2} \left[x_\uA^2 \xi_\uA^2 + \frac{F^2 \sigma_\uM^2 \left(1 - x_\uA^2 \right)^2 \sin^2(\theta_\uA)}{x_\uA^2 \cos^2(\theta_\uA + \psi_\uA)} \right]
\label{mu2}
\end{equation}
The Alfv\'en regularity condition is obtained by substituting these relations into the transfield equation.
\begin{align}
\frac{F^2 \sigma_\uM^2 (1 - x_\uA^2) (\sigma_\uA + 1)^2 \sin(\theta_\uA)}{\mu^2 \cos^2(\theta_\uA + \psi_\uA)} \bigg\{ -2 \frac{\Gamma - 1}{\Gamma} \frac{(F - 2) ( \xi_\uA - 1) (1 - x_\uA^2)}{x_\uA^2 \xi_\uA} \sin(\theta_\uA) + 2 \cos(\psi_\uA) \sin(\theta_\uA + \psi_\uA) \frac{\sigma_\uA + 1}{\sigma_\uA} \nonumber \\
+ \frac{\sin(\theta_\uA)}{x_\uA^2} \left[ (F - 1) (1 - x_\uA^2) - 1\right] \bigg\} \nonumber \\
= \left[ x_\uA^2 - \sigma_\uA (1 - x_\uA^2) \right]^2 - (F - 1) \sigma_\uA^2 (1 - x_\uA^2) - 2 \frac{\Gamma - 1}{\Gamma} (F - 2) \frac{\xi_\uA - 1}{\xi_\uA} \left\{ x_\uA^2 - \left[ x_\uA^2 - \sigma_\uA (1 - x_\uA^2) \right]^2 \right\}
\label{ARC}
\end{align}

The wind equation is given by
\begin{subequations}
\label{windequation}
\begin{equation}
\frac{\ud M^2}{\ud \theta} = \frac{C_1 B_2 - C_2 B_1}{A_1 B_2 - A_2 B_1}
\end{equation}
where
\begin{align}
A_1 &= - 2 x^4 \cos^2(\psi + \theta) (1 - M^2 - x^2) \Bigg\{ \mu^2 \frac{x_\uA^2 M^2 (1 - G^2)^2}{(1 - M^2 - x^2)^2} \nonumber \\
&\qquad + G^2 (1 - M^2 - x^2) \left( \frac{F^2 \sigma_\uM^2 M^2 \sin^2(\theta)}{x^4 \cos^2(\psi + \theta)} - \frac{\xi^{(\Gamma + 1)} q^{\Gamma - 1}}{M^{2 \Gamma} \left[ \xi \left( \frac{\Gamma - 2}{\Gamma - 1} \right) + 1 \right] } \right) \Bigg\} \\
B_1 &= - 2 F^2 \sigma_\uM^2 M^4 \sin^2(\theta) \tan(\psi + \theta) G^2 (1 - M^2 - x^2)^2 \\
C_1 &= x^4 \cos^2(\psi + \theta) \mu^2 \frac{\ud G^2}{\ud \theta} \left[ - \left( 1 - M^2 - x_\uA^2 \right)^2 + 2 x_\uA^2 \left( G^2 - M^2 -x^2 \right) (1 - x_\uA^2) \right] \nonumber \\
&\qquad + \frac{\ud G^2}{\ud \theta} (1 - M^2 -x^2) \left( 1- M^2 - 3 x^2 \right) \left[ \xi^2 x^4 \cos^2(\psi + \theta) + F^2 \sigma_\uM^2 M^4 \sin^2(\theta) \right] \nonumber \\
&\qquad + 2 F^2 \sigma_\uM^2 M^4 \sin^2(\theta) G^2 (1 - M^2 - x^2)^2 \left[ \frac{\cos(\theta)}{\sin(\theta)} - \frac{1}{G^2} \frac{\ud G^2}{\ud \theta} + \tan(\psi + \theta) \right] \\
A_2 &= -\sin^2(\theta) \tan(\psi + \theta) \\
B_2 &= \sin^2(\theta)\frac{1 - M^2 - x^2}{\cos^2(\psi + \theta)} \\
C_2 &= -\sin^2(\theta)\frac{1- M^2 - x^2}{\cos^2(\psi + \theta)} + \sin^2(\theta) \tan(\psi + \theta) \left[ x_\uA^2 \frac{\ud G^2}{\ud \theta} + (1 - M^2 - x^2) \frac{1}{G} \frac{\ud G}{\ud \theta} \right] \nonumber \\
&\qquad + (F-1) \frac{x_\uA^4 \mu^2 x^2}{F^2 \sigma_\uM^2} \left( \frac{1 - G^2}{1 - M^2 - x^2} \right)^2 - \sin^2(\theta) \frac{M^2 + F x^2 - F + 1}{\cos^2(\psi + \theta)} \nonumber \\
&\qquad - \frac{x_\uA^4 \mu^2 x^2}{F^2 \sigma_\uM^2 M^2} \left( \frac{G^2 - M^2 - x^2}{1 - M^2 - x^2} \right)^2 + 2 \frac{\Gamma - 1}{\Gamma} \frac{F - 2}{F^2 \sigma_\uM^2} \frac{\xi (\xi - 1) x^4}{M^2}
\end{align}
\end{subequations}

\end{document}